\documentclass[11pt]{article}

\usepackage{amsmath,amssymb,amsfonts,amsthm} 
\usepackage{mathrsfs,latexsym} 
\usepackage{mathtools}

\usepackage[mathscr]{eucal} 

\usepackage{color}
\usepackage{cite}
\usepackage{epsfig}
\usepackage{graphicx}

\usepackage{bm}
\usepackage{cite}

\numberwithin{equation}{section}

\newcommand{\RR}{{\mathbb R}}

\newcommand{\Cc}{{\mathcal{C}}}
\newcommand{\Dc}{{\mathcal{D}}}

\newcommand{\Lc}{{\mathcal{L}}}

\newcommand{\Oc}{{\mathcal{O}}}

\newcommand{\bdelta}{{\bm \delta}}

\newcommand{\bDelta}{\bm \Delta} 
\newcommand{\bnabla}{\bm \nabla}

\newcommand{\bfi}{{\bm f}}
\newcommand{\bgi}{{\bm g}}

\newcommand{\bmi}{{\bm m}}
\newcommand{\bni}{{\bm n}}
\newcommand{\bpi}{{\bm p}}

\newcommand{\bxi}{{\bm x}}

\newcommand{\bAi}{{\bm A}}
\newcommand{\bBi}{{\bm B}}

\newcommand{\bEi}{{\bm E}}

\newcommand{\Vf}{{\mathfrak V}}
\newcommand{\Wf}{{\mathfrak W}}

\newcommand{\Zf}{{\mathfrak Z}}

\newcommand{\supp}{{\mathrm{supp}}}

\newcommand{\ad}[1]{\mbox{Ad}(#1)}

\def\ie{{\it i.e.\ }}
\def\viz{{\it viz.\ }}

\newcommand{\be}{\begin{equation}}
\newcommand{\ee}{\end{equation}}

\begin{document} 

%

\title{\LARGE Gauss's law, the manifestations of gauge fields, \\ and their  
  impact on local  observables \\[4mm]
\large  Dedicated to the memory of Giovanni Morchio}
\author{Detlev Buchholz${}^a$,
Fabio Ciolli${}^b$, Giuseppe Ruzzi${}^c$ and   
Ezio Vasselli${}^c$ \\[10pt]
\small  
${}^a$ Mathematisches Institut, Universit\"at G\"ottingen, \\
\small Bunsenstr.\ 3-5, 37073 G\"ottingen, Germany\\[5pt]
\small
${}^b$ Dipartimento di Matematica e Informatica, Universit\'a della Calabria \\
\small Via Pietro Bucci - Cubo 30B, 87036 Rende (CS), Italy \\[5pt] 
\small
${}^c$ 
Dipartimento di Matematica, Universit\'a di Roma Tor Vergata \\
\small Via della Ricerca Scientifica 1, 00133 Roma, Italy \\
}
\date{\today}

\maketitle

\noindent \textbf{Abstract.}   
Within the framework of the universal algebra of the 
electromagnetic field, the impact of globally neutral configurations
of external charges on the field is analyzed.
External charges are not affected by the field,
but they induce localized automorphisms of the universal  
algebra. Gauss's law implies that these
automorphisms cannot be implemented by 
unitary operators involving only the electromagnetic field,
they are outer automorphisms. The missing degrees 
of freedom can be incorporated in an enlargement of the universal
algebra, which can concretely be represented by exponential
functions of gauge fields and an abelian algebra describing the
external charges. In this manner, gauge fields manifest themselves
in the framework of gauge invariant observables.  
The action of the automorphisms
on the vacuum state gives rise to representations
of the electromagnetic field with vanishing global charge, 
which are locally disjoint from the vacuum
representation. This feature disappears in the enlarged 
universal algebra of the electromagnetic field.
The energy content of the states is 
well defined in both cases and bounded from below.
The passage from these globally   
neutral states to charged states and the determination of their 
energy content are also being discussed. 



\section{Introduction}
\label{sec1} 
\setcounter{equation}{0} 

Gauss's law, relating the electric charge to the
surrounding flux of the electromagnetic field, is
the most distinctive feature of quantum electrodynamics.
Its numerous implications for the structure of the theory have been 
widely discussed in the literature, most thoroughly by G.\ Morchio
and his longtime scientific companion F.\ Strocchi,
cf.\ \cite{MoSt1983} and references quoted there. One of the 
fundamental insights gained in these studies is the observation that
field operators, creating electrically charged states
as well as their Coulomb fields  
from a vacuum state, cannot be localized in compact spacetime
regions. Charged states are limits of neutral states, involving pairs   
of opposite charges, where one of the charges is shifted to spacelike
infinity. In contrast to massive theories, there remains
in this limit a Coulomb field as an inevitable memory effect that requires
a description by non-local operators. 

\medskip 
These features are commonly attributed to the fact that
quantum electrodynamics is a
gauge theory. Yet gauge fields do
not have a direct physical significance. So one had to clarify 
on one hand how these fields can be eliminated in computations
of the physical observables of the theory.
On the other hand, it raised the question whether
the usage of gauge fields is really necessary for the formulation of the
theory, or whether it is just a convenient calculational tool
without further physical significance, cf.\ for example
the novel approach to quantum electrodynamics in \cite{MuReSch}. 

\medskip
Morchio and Strocchi commented on this issue by a 
thoughtful closing remark in Ref.\ \cite{MoSt2007}. They wrote:
``The validity of local Gauss' laws appears to have a more direct physical
meaning than the gauge symmetry, which is non-trivial only on non-observable
fields. It is therefore tempting to regard the validity of local Gauss'
laws as the basic characteristic feature of gauge field theories, and to
consider gauge invariance merely as a useful recipe for writing down
Lagrangian functions which automatically lead to the validity of local
Gauss' laws.''  

\medskip
It is the aim of the present article to shed further light on this
issue. In order to understand whether gauge fields are an
indispensible  part of the theory, one must proceed
from observables, while avoiding 
\textit{a priori} assumptions about non-observable
structures. In the case at hand, the basic observable ingredient is the 
electromagnetic field, satisfying the homogeneous Maxwell equation. 
It is related to the electric  current by the
divergence of its Hodge dual. Being an observable, it has also to
comply with the condition of locality (Einstein causality).
It has been shown in \cite{BuCiRuVa2015} that these features 
can consistently be incorporated in a universal C*-algebra
of the electromagnetic field that underlies every theory
of electromagnetism. This general framework provides the basis 
for our analysis of the problem at hand.  

\medskip 
To analyze the effects of Gauss's law on the structure
of the theory, one must first understand whether matter 
which carries an electric charge can be sufficiently well localized,
as opposed to the infinitely extended
Coulomb-like field that emanates 
from it. In a long-term project, Morchio, along with four close 
colleagues, has been working on this problem. They showed that a 
sharp localization in compact spacetime regions
is possible in the classical Maxwell-Dirac theory and also in some
non-interacting quantized models
\cite{BuDoMoRoSt97,BuDoMoRoSt01}.
It turned out, however, that realistic quantized and electrically
charged matter cannot be localized in such a manner \cite{BuDoMoRoSt01}.
This led to a new, but rather heavy general framework covering 
electrically charged systems \cite{BuDoMoRoSt07}. 

\medskip
These complications are not necessary, however,  
to clarify the role of gauge fields.
We can rely here on the idealization that the matter, carrying
an electric charge, is of an external nature (\ie it is not
affected by interactions with the electromagnetic field).
As has been shown in \cite{BuCiRuVa2021}, one can then describe
the impact of the electric charges on the electromagnetic field
by automorphisms of the corresponding universal algebra.
In case of pairs of opposite charges located in compact spacetime
regions, the corresponding automorphisms are well-localized.
Even though their global charge is zero, they
cannot be unitarily implemented, however, by operators involving only the
electromagnetic field; they are outer automorphisms.
This shows that the dipole field, connecting the charges, comprises 
some additional degrees of freedom, indicating the presence of
gauge fields. In fact, the automorphisms can be implemented by
exponential functions of gauge fields 
in the Gupta-Bleuler framework, which thus generate the
dipole field. The accompanying
charged matter can be described
by an abelian algebra of local fields which commute
with the electromagnetic field, but have non-trivial
commutation relations with the gauge charges.
The resulting algebra of gauge and matter fields
is an improved version  
of a preliminary proposal made in \cite{BuCiRuVa2021}.

\medskip
It is the aim of the present article to elaborate on these
observations in detail. In particular, we will analyze
the properties of the states which are obtained by composing
the basic (charge-free) vacuum state on the universal
algebra with the localized charge-containing automorphisms. The
resulting states give rise to representations which are disjoint 
(even locally) from the basic vacuum representation, in spite of the
fact that they have a vanishing global charge. This apparent conflict with the
Doplicher-Haag-Roberts approach to sector analysis \cite{DoHaRo} 
resolves if one recalls that the local subalgebras
of the universal algebra are not weakly closed, \ie they are not
factors of type~III, as is frequently taken for granted in
algebraic quantum field theory. In fact,
these algebras contain a primitive two-sided ideal. It is generated by
the current, which vanishes in the basic vacuum
representation but becomes non-trivial in the presence of
genuine charge distributions.

\medskip
To remedy this undesirable feature, we enlarge 
the algebra by adding operators which are complementary to
the current. It leads again to a version of the universal algebra. 
The charge-containing states on this algebra are then locally normal
relative to each other.
Despite these differences, the states have in both cases 
a well-defined energy content that is bounded
from below. We will also discuss how states with non-vanishing global
charge arise as limits of these globally neutral states and determine
their energy content.

\medskip
Models of the latter type have been previously discussed in the
literature. There gauge and matter fields are generally taken
as input, cf.\ for examp\-le \cite{He,Mo}; we do not add much to that issue.
The principal subject of our investigation is the longstanding 
question of whether the presence of gauge fields and their
physical significance can
be uncovered from the local observables \cite{Ha}.
An affirmative answer would corroborate the view that the physically 
relevant information of a theory
is encoded in the observables and can be extracted
from them. Such an analysis has been exceedingly successful in case of
fields carrying a global gauge charge in massive theories
\cite{DoHaRo,DoRo}. However, similar convincing results have not been
obtained up to now in case of local gauge fields. The present results
are a modest step, pointing into the direction of an affirmative
answer. Yet they are far from giving a complete solution.

\medskip
Our article is organized as follows. In the subsequent
Sect.\ 2 we recall the definition of the universal algebra
of the electromagnetic field, presented in \cite{BuCiRuVa2015}, 
and improve on the construction of local, 
charge-containing automorphisms and their unitary implementers,
proposed in \cite{BuCiRuVa2021}.
In Sect.\ 3, being the central part of our article,
we illustrate the abstract framework by various   
concrete examples, based on the Gupta-Bleuler
formalism. Within this setting, we discuss the properties
and the relations between
the resulting charge-containing representations of the
universal algebra. We establish the existence 
of meaningful dynamics and determine the energetic properties of
the charge-containing states. In Sect.\ 4 we discuss
how charged states are obtained as limits of these globally neutral
states and we summarize our findings in the conclusions. 

\section{Universal algebra and external charges}
\label{sec2}
\setcounter{equation}{0} 

For the convenience of the reader, we recall in this section the
definition and basic properties of the universal  
C*-algebra $\Vf$ of the electromagnetic field,
introduced in \cite{BuCiRuVa2015}. We outline 
how local charge measurements are described in this setting  
and improve on the construction of localized outer automorphisms,
considered in \cite{BuCiRuVa2021}. These automorphisms describe the
impact of external charges on the electromagnetic field and 
are shown to be unitarily implemented by the action of local gauge and
matter fields. 

\medskip 
In heuristic terms, the universal algebra is generated by 
exponential functions of the electromagnetic field $F$, 
which can conveniently be represented by an intrinsic (gauge
invariant) vector potential $A_I$, \viz $e^{i F(f)} = e^{i A_I(g)}$.
Here $f \in \Dc_2(\RR^4)$ is any real, 
skew-tensor-valued test function with compact support.
The corresponding function $g$ is a solution 
of the equation $g = \delta f$, where $\delta = \star d \star$
is the exterior co-derivative, $\star$ denotes the
Hodge operator, and $d$ is the exterior
derivative. The space of the real, vector-valued test functions
$g$, satisfying $\delta g = 0$, is denoted by $\Cc_1(\RR^4)$.
Since $F$ satisfies the homogeneous Maxwell equation, the
potential $A_I(g)$ is unambiguously defined for any
given $F(f)$ by Poincar\'e's Lemma.

\medskip
The exponentials
of the intrinsic vector potential are described by 
unitary operators $V(a g)$, where $a \in \RR$ and
$g \in \Cc_1(\RR^4)$; they 
generate a *-algebra~$\Vf_0$. These operators  are  
subject to relations, expressing basic algebraic and
locality properties of the potential, where 
$g_\cdot \in \Cc_1(\RR^4)$, $\, f_\cdot \in \Dc_2(\RR^4)$, 
\begin{eqnarray} \label{e.2.1}
& V(a_1 g) V(a_2 g) = V((a_1 + a_2) g) \, , \ 
V(g)^* = V(-g) \, , \ 
V(0) = {\mathbf 1} \, , \ \ \\ 
\label{e.2.2}
& V(\delta f_1) V(\delta f_2) =
V(\delta f_1 + \delta f_2)
 \ \ \text{if} \ \  
\supp \, f_1 \perp \supp \, f_2 \, , & \\
\label{e.2.3}
& \lfloor V(g_1) , V(g_2) \rfloor \, \in \Vf_0 \cap \Vf_0'
\ \ \text{if} \ \ 
\supp \, g_1  \perp  \supp \, g_2 
\, . &
\end{eqnarray}
Here the symbol $\perp$ between two regions indicates that they are
spacelike separated,  $\Vf_0 \cap \Vf_0'$ is the center of $\Vf_0$, 
and 
$\lfloor X_1, X_2 \rfloor \coloneqq X_1 X_2 X_1^* X_2^*$ denotes the
group theoretic commutator of $X_1, X_2$. We refer to
\cite{BuCiRuVa2015, BuCiRuVa2021} for a discussion of the
physical significance of these relations. 

\medskip
On the algebra $\Vf_0$ there exists a particular faithful state 
which has all unitaries in its kernel, apart from $\mathbf 1$ 
(counting measure). 
The corresponding GNS-representation determines
a C*-norm on this algebra. Proceeding to the
completion of $\Vf_0$ with regard to its maximal C*-norm,
resulting from the set of all of its states, 
one arrives at the C*-algebra $\Vf$, the universal algebra
of the electromagnetic field \cite{BuCiRuVa2015}. This algebra admits an
automorphic action of the proper orthochronous Poincar\'e
group which is fixed by the relations
\be \label{e.2.4} 
\alpha_P(V(g)) \coloneqq V(g_P)  \, , \quad P \in
\Lc_+^\uparrow \ltimes \RR^4 \, , \ g \in \Cc_1(\RR^4) \, ,
\ee
where $x \mapsto g_P{}^\mu(x) \coloneqq L^\mu{}_\nu \, g^\nu(L^{-1}(x - y))
\in \Cc_1(\RR^4)$ \ for \ $P = (L,y)$. 

\medskip
Whereas the algebra $\Vf$ does not include elements which create
electric charges, it contains all ingredients for their analysis. 
To recall this fact, we resort to the heuristic picture
(being meaningful in regular states) that 
the underlying unitaries are exponential functions of the 
electromagnetic field $F$. It determines the electric current $J$
by the inhomogeneous Maxwell equation,
\be \label{e.2.5} 
J(h) \coloneqq F(dh) = A_I(\delta d h) \, , \quad h \in \Dc_1(\RR^4) \, .
\ee
Here $\Dc_1(\RR^4)$ is the space of real, vector-valued test functions
with compact support. The zero component of the current determines
local charge operators for suitable choices of the test functions $h$.
Its expectation values can be changed by linear maps of the 
intrinsic vector potential of the form 
\be \label{e.2.6}
A_I(g) \mapsto A_I(g) + \varphi(g) {\mathbf 1} \, ,
\quad g \in \Cc_1(\RR^4) \, ,
\ee
where $\varphi : \Cc_1(\RR^4) \rightarrow \RR$ is any real
linear functional. 
Applying these maps to the current, one obtains
\be \label{e.2.7}
J(h) \mapsto J(h) + \varphi(\delta d h)
{\mathbf 1} \, , \quad h \in \Dc_1(\RR^4) \, .
\ee

The choice of functions $h \in \Dc_1(\RR^4)$, corresponding
to charge measurements
in a given spacetime region, and of functionals
$\varphi: \Cc_1(\RR^4) \rightarrow \RR$, describing
configurations of external charges, were discussed in detail
in \cite{BuCiRuVa2021}, cf.\ also Sect.~4.
Choosing a Lorentz frame,
charge measurements are described by test functions $h \in \Dc_1(\RR^4)$
with vanishing spatial components and time components of the form 
$x \mapsto \tau(x_0) \chi(\bxi)$,
where $\tau$ and $\chi$ fix the time
and spatial region where charges are to be determined. The
resulting test functions for the intrinsic vector
potential are given by
\be \label{e.2.8} 
x \mapsto \delta d h(x) =
(\tau(x_0) \bDelta \chi(\bxi), \, \dot{\tau}(x_0)
\bnabla \chi(\bxi)) \, .
\ee
Here $\bnabla$ denotes the spatial gradient, $\bDelta$ the Laplacian,
and the dot $\dot{ }$ indicates a time derivative.

\medskip
The functionals of interest here have the form
\be \label{e.2.9}
\varphi_m(g) \coloneqq - 
\int \! dx dy \ m^\mu(x) \, D(x-y) \, g_\mu(y) \, ,
\quad g \in \Cc_1(\RR^4) \, ,
\ee
where $m \in \Dc_1(\RR^4)$ and $D$ denotes the zero mass Pauli-Jordan
commutator function. Note that they vanish if $m$ and $g$ have
spacelike separated supports. 
The co-derivative $\delta m$ can be interpreted as 
a density of external charges, which are not affected by the
electromagnetic field, and $\varphi_m$ describes their impact on this 
field. (We restrict ourselves to 
well-behaved functions $m$ for the sake of simplicity, but
signed measures can be admitted.)  Since $m$ has compact support, the 
global charge determined by $\varphi_m$ is zero. Yet 
if the support of $\delta m$ consists of spacelike
separated, compact regions, each of them may contain a charge
which is different from zero.
The external charge content in these subregions can be precisely determined
by means of the local charge operators, defined above, and Gauss's
law, cf.\ \cite{BuCiRuVa2021} and Sect.~4.

\medskip
These observations can be transferred to the
algebraic framework. The maps $\varphi_m$,
defined in relation \eqref{e.2.9}, determine automorphisms
$\beta_m$ of the
universal algebra~$\Vf$ which act on the generating unitaries,
\ie the exponential functions of the electromagnetic field,
according~to 
\be \label{e.2.10} 
\beta_m(V(g)) \coloneqq
e^{i \varphi_m(g)} \, V(g) \, ,
\quad g \in \Cc_1(\RR^4) \, .
\ee
The composition of these automorphisms with 
Poincar\'e transformations~$P$ satisfies 
$\alpha_P \beta_m = \beta_{m_P} \alpha_P$,
where $m_P(g) = m(g_{P^{-1}})$ for $g \in \Cc_1(\RR^4)$. 

\medskip 
Identifying the 
exponential functions of the local charge operators $J(h)$ 
with the unitaries $V(\delta d h)$, where $\delta d h$ is
given by equation \eqref{e.2.8}, it follows that
\be \label{e.2.11}
\beta_m(V(\delta d h)) =  e^{i \varphi_m(\delta d h)} \,  V(\delta d h) \, , 
\quad h \in \Dc_1(\RR^4) \, .
\ee
It was shown in \cite{BuCiRuVa2021} that for functionals
$\varphi_m$, based on non-trivial charge distributions $\delta m$,  
and for
suitable test functions $h$, the phase factors in equation~\eqref{e.2.11}
are different from $1$. In these cases the maps $\beta_m$
define outer automorphisms of the universal algebra $\Vf$. The latter 
assertion follows from the existence of a  vacuum
representation of $\Vf$, describing the non-interacting electromagnetic
field \cite{BuCiRuVa2015}. There all unitaries
$V(\delta d h)$ are represented by~${\mathbf 1}$, 
whence all local charge operators are
equal to zero.  This fact excludes the existence of
unitary operators in the universal algebra $\Vf$ which implement the
action of $\beta_m$ in equation \eqref{e.2.11}. In order to implement 
it, one must extend the algebra $\Vf$, which is
accomplished as follows. 

\medskip
One considers a family of operators 
$W(m)$, $m \in \Dc_1(\RR^4)$, satisfying for 
$a_1, a_2 \in \RR$ and $m, m_1, m_2 \in \Dc_1(\RR^4)$ the
equalities 
\begin{eqnarray} \label{e.2.12}
& W(a_1 m) W(a_2 m) = W((a_1 + a_2) m) \, , \, 
W(m)^* = W(-m) \, , \, 
W(0) = {\mathbf 1} \, , \qquad \\ 
\label{e.2.13}
& W(m_1) W(m_2) = W(m_1 + m_2)  \ \ \text{if} \ \  
\supp \, m_1 \perp \supp \, m_2 \, .  
\end{eqnarray}
These relations encode the information that each $W(m)$ 
is a unitary operator with causal localization properties, determined by
the support of $m$. One can also define 
Poincar\'e transformations,  putting
$\alpha_P(W(m)) = W(m_P)$, where $m_P$ is defined similarly to  
equation \eqref{e.2.4}.
The assumption that the unitaries $W(m)$ induce the automorphisms
$\beta_m$ is encoded in the equalities 
\be \label{e.2.14} 
W(m) V = \beta_{m}(V) W(m) \, ,  \quad  V \in \Vf  \, ,
 \ m \in \Dc(\RR^4) \, . 
 \ee
One then proceeds from the unitary groups generated by
$V(g)$, $g \in \Cc_1(\RR^4)$, respectively $W(m)$, $m \in \Dc_1(\RR^4)$,
to their semi-direct product, fixed by equation~\eqref{e.2.14}.
By a similar procedure as in case of the universal algebra
(existence of a counting measure), one
obtains a C*-algebra $\Wf \supset \Vf$. Note that its 
elements $W(m)$ only describe the effects of external
charges. In addition, there could be dynamical charges present
which would manifest themselves by further relations within the universal
algebra of the electromagnetic field, depending on
details of the dynamics. 

\medskip
It turns out that the unitaries $W(m)$, implementing the
automorphisms $\beta_m$, are not fixed by equation \eqref{e.2.14}.
There exists an
abundance of operators inducing the same action. It is an indication
that the charged system bears additional local degrees of freedom.
From our present point of view, this fact manifests itself by
the existence of local gauge transformations. 
Picking any scalar distribution $s$ on $\RR^4$, these 
transformations are given by
\be \label{e.2.15} 
\gamma_s(W(m)) \coloneqq
e^{i \int \! dx \, (ds)_\mu(x) \, m^\mu(x)} \, W(m)
= e^{- i \int \! dx \, s(x) \, \delta m(x)} \, W(m) \, .
\ee
So $\gamma_s$ acts trivially on $W(m)$ if $\delta m = 0$, and this
triviality of action holds similarly for all elements of $\Vf$.

\medskip
These in the formalism still missing
degrees of freedom are described by
charged matter fields, which compensate the gauge charges
carried by the operators $W(m)$. 
Their combined action then defines a unique 
(\ie gauge-invariant) local operation, inducing
automorphisms creating the charge distributions.
In the present simple case of external charges,
the charged matter fields can be described by
the elements of an abelian algebra.
Given any real, scalar test functions 
$\rho \in \Dc_0(\RR^4)$, this algebra is generated by 
unitary operators $\psi(\rho)$, the matter fields, which
are subject to the relations 
\be \label{e.2.16} 
\psi(\rho_1) \psi(\rho_2) = \psi(\rho_1 + \rho_2) \, ,
\quad \psi(\rho)^* = \psi(-\rho) \, , 
\quad \psi(0) = {\mathbf 1} \, .
\ee
The Poincar\'e transformations $P$ act on the fields by 
$\alpha_P(\psi(\rho)) = \psi(\rho_P)$ and, most importantly,  
they transform under gauge transformations according to
\be \label{e.2.17}
\gamma_s(\psi(\rho)) = e^{i \int \! dx \, s(x) \rho(x)} \psi(\rho) \, . 
\ee
Thus the operators $\psi(\delta m) W(m)$ and 
$W(m) \psi(\delta m)$ are gauge invariant and
we assume that they are equal. Yet, in contrast to the discussion in
\cite{BuCiRuVa2021}, we do not assume from the outset that
the matter fields commute with all elements of $\Wf$.
We only require that they commute with the electromagnetic field,
\be  \label{e.2.18}
\psi(\rho) V = V \psi(\rho) \, , \qquad \rho \in \Dc_0(\RR^4) \, , \
V \in \Vf \, .
\ee
The physical picture behind this assumption is the idea that the
impact of the external charged matter and its electromagnetic
tail on the electromagnetic field is fully described by the
automorphisms $\beta_m$
of the universal algebra $\Vf$. There are no other interactions
between the external  matter and the electromagnetic field. 

\medskip
The relations given
above define a consistent extension of the universal algebra $\Vf$
by gauge and matter fields.
This becomes apparent if one notices that all relations involve
unitary operators, \ie they determine a unitary group.
One can then proceed to a corresponding C*-algebra 
by the same token as in case of the universal algebra
$\Vf$, \ie one makes use again of the existence of a faithful state (counting
measure). The resulting algebra has a local net structure, fixed by the
supports of the underlying test functions, Poincar\'e transformations
act on it covariantly, and the algebra is stable under the
automorphic action of gauge transformations. We refrain from 
proving these statements here. Instead, we present in the subsequent
section a concrete representation of the present abstract algebraic 
framework. 

\section{Neutral states with varying charge densities}
\label{sec3}
\setcounter{equation}{0}

We construct in this section concrete
representations of the abstract universal algebra 
that describe external charge distributions with vanishing global charge. 
As in \cite{BuCiRuVa2021}, we make use of the Gupta-Bleuler
formalism. Since this setting does not fix a gauge, \ie  
incorporates operators that generate
non-trivial gauge-transformations, we must also specify
commutation relations between the Gupta-Bleuler fields 
and the matter fields. This step was missing in \cite{BuCiRuVa2021}. 

\medskip
We begin by briefly recalling  the Gupta-Bleuler
framework, cf.\ \cite{St,Str}.  The exponentials of the
Gupta-Bleuler fields are denoted by the symbols 
\be \label{e.3.1}
e^{iA(f)} \, , \quad f \in \Dc_1(\RR^4) \, .
\ee
They satisfy the equations for $a \in \RR$ and $f,g \in \Dc_1(\RR^4)$ 
\be \label{e.3.2} 
e^{iA(f)} e^{iaA(g)} = e^{-i (a/2) \langle f, D g \rangle} \, e^{iA(f + a g)} ,
\ \, \big( e^{iA(f)} \big) {}^* =  e^{- iA(f)} ,
\ \,  e^{iA(0)} = {\mathbf 1} \, ,
\ee
where we made use of the shorthand notation 
\be \label{e.3.3} 
\langle f, D g \rangle \coloneqq \int \! dx dy \, f_\mu(x) \, 
D(x-y) \, g^\mu(y) \, .
\ee
Note that we do not assume that the
Gupta-Bleuler fields are solutions of
the wave equation. However, since $D$ is a bi-solution of the
wave equation, the operators~$e^{iA(\square f)}$, where
$\square$ is the d'Alembertian, commute with all
other operators, \ie they are central elements of the Weyl algebra
generated by the Gupta-Bleuler fields. 

\medskip
We represent now the abstract operators, defined in the preceding section, 
by these fields and deal with them in the remainder of
this article. With this understanding we put, by some slight abuse of
notation, 
\be \label{e.3.4} 
W(m) \coloneqq e^{i A(m)} \, , \quad m \in \Dc_1(\RR^4) \, .
\ee
Assuming that there are only external charges present, we can
identify the exponentials for functions in 
the subspace $\Cc_1(\RR^4) \subset \Dc_1(\RR^4)$
with the generating elements of the universal algebra $\Vf$,
\be \label{e.3.5} 
V(g) \coloneqq e^{i A(g)} = W(g)  \, , \quad g \in \Cc_1(\RR^4) \, .
\ee
In view of the underlying equations \eqref{e.3.2},
the latter unitaries comply with relations \eqref{e.2.1} to \eqref{e.2.3}.
Note that the exponentials of the current $V(\delta dh)$, cf.\ equation
\eqref{e.2.5}, are elements of the center of $\Vf$. The former
unitaries $W(m)$ satisfy equations \eqref{e.2.12}, \eqref{e.2.13}.
They generate the algebra $\Wf$ in the present
setting and induce the action of the
automorphisms $\beta_m$, in accordance with equation~\eqref{e.2.14}.

\medskip
The algebraic properties of the Gupta-Bleuler fields also imply that 
for $s \in \Dc_0(\RR^4)$ and $m \in \Dc_1(\RR^4)$ one has 
\be \label{e.3.6} 
W(ds) W(m) W(ds)^* = e^{- i \int \! dx \, (Ds)(x) \, 
  (\delta m)(x)} \, W(m) \, .
\ee
Hence the operators $W(ds)$ induce non-trivial
gauge transformations of the unitaries $W(m)$ if $\delta m \neq 0$.
They amount to
shifts of the underlying potential by $d Ds$. Note that the
function $x \mapsto (Ds)(x)$, obtained by convolution of 
$D$ with $s$, is a solution of the wave equation, whence 
not a test function. 

\medskip 
The matter field $\psi(\rho)$, $\rho \in \Dc_0(\RR^4)$,
introduced in the preceding section,
was assumed to commute with the electromagnetic field
but not with gauge fields. This requires to postulate in the
present setting specific commutation relations between the
matter fields and the Gupta-Bleuler fields. There  
we rely on the fact that the Pauli-Jordan distribution $D$ can uniquely
be split into a retarded and an advanced part, $D = D_r - D_a$,
which are Riesz distributions, cf.~\cite[Sect.~1.2]{BaGiPf}.
Their products $D_r^2 \coloneqq D_r * D_r$ and
\mbox{$D_a^2 \coloneqq D_a * D_a$}, defined by convolution,
also belong to this class. Making use of this fact, we impose
for $m \in \Dc_1(\RR^4)$ and $\rho \in \Dc_0(\RR^4)$  the relations  
\be \label{e.3.7}  
W(m) \, \psi(\rho) \,  W(m)^* = 
e^{-i\langle \delta m, D^{\hspace*{0.6pt} \flat} \rho \rangle } \, \psi(\rho) \, , 
\ee
where 
\be \label{e.3.8}
\langle \delta m, D^{\hspace*{0.8pt} \flat } \rho \rangle
\coloneqq \int \! dx dy \, (\delta m) (x) \,
(D_r^2(x-y) - D_a^2(x - y)) \, \rho(y) \, . 
\ee
Thus $\psi(\rho)$ commutes with $W(m)$ if 
$\delta m = 0$. Since
$\langle \delta m, D^{\hspace*{0.8pt} \flat} \rho \rangle
= - \langle \rho, D^{\hspace*{0.8pt} \flat} \delta m \rangle$,
it is also clear that $\psi(\delta m)$ and $W(m)$ commute
for all $m \in \Dc_1(\RR^4)$. Finally, if $m = d s$ with 
$s \in \Dc_0(\RR^4)$, whence $\delta m = \square s$,
one has
$\langle \delta m, D^{\hspace*{0.8pt} \flat } \rho \rangle
= \langle s, \square D^{\hspace*{0.8pt} \flat } \rho \rangle
$. The relation $\square \, D^\flat = D$,
cf.\ \cite[Prop.\ 1.2.4]{BaGiPf}, then implies  
\be \label{e.3.9} 
W(ds) \psi(\rho) W(ds)^* = e^{ i \int \! dx \, (Ds)(x) \rho(x)} \,  \psi(\rho) \, .
\ee
Hence the unitaries $W(ds)$ induce on the matter fields the
gauge transformations $Ds$, in accordance with their action
on the potentials in \eqref{e.3.6}. 
In particular, the operators
$\psi(\delta m) W(m) = W(m) \psi(\delta m)$ 
are gauge invariant for any choice of $m \in \Dc_1(\RR^4)$. 
So, within the present framework, the matter field 
has all properties postulated in the preceding section. 
Moreover, since~$D^{\hspace*{0.8pt} \flat}$ is a causal distribution,
the operators $\psi(\delta m) W(m)$ are local, \ie their
commutators vanish for test functions $m$ with
spacelike separated supports. 

\medskip
We denote by $\Zf$ the algebra that is generated by
the gauge invariant operators
$\psi(\delta m) W(m)$, $m \in \Dc_1(\RR^4)$.
It is an extension of the algebra $\Vf$, which is generated 
by the restriction of these operators to the subspace of functions $m$
satisfying $\delta m = 0$. The preceding
relations and the algebraic properties of the matter fields
imply, $m, m_1, m_2 \in \Dc_1(\RR^4)$, 
\begin{align} \label{e.3.10}
& \psi(\delta m_1) W(m_1) \,  \psi(\delta m_2) W(m_2) 
= \eta \, \psi(\delta(m_1 + m_2)) W(m_1 + m_2)  \nonumber \\ 
& (\psi(\delta m) W(m))^*  = W(-m) \psi(- \delta m)
= \psi(- \delta m)  W(-m) \, ,
\end{align}
where $\eta$ are phase factors.
It follows that the linear span of these gauge invariant operators is
norm dense in $\Zf$.

\medskip 
We rearrange their sums into groups
of operators creating the same charge
density: any pair of functions $m_1, m_2 \in \Dc_1(\RR^4)$ satisfying 
$\delta m_1 = \delta m_2$ differs by some element of $\Cc_1(\RR^4)$.
Thus every sum of gauge invariant operators
$\sum_j  c_j \, \psi(\delta m_j)  W(m_j)$, where 
all charge densities coincide, $\delta m_j = \delta m$,
can be presented with the help of relation \eqref{e.3.2}
in the standard form
\be \label{e.3.11} 
S(\delta m) \coloneqq \psi(\delta m)  W(m) \sum_i c_i \eta_i  V(g_{i}) \, ,
\ g_{i} \in  \Cc_1(\RR^4) \, .
\ee
An arbitrary sum of gauge invariant operators can be presented as 
a sum $\sum_j S(\delta m_j)$ of operators $S(\delta m_j)$ with 
different charge densities $\delta m_j$.

\medskip
We are now in the position to exhibit a  
Poincar\'e invariant state $\omega_0$ on $\Zf$, which  
extends the non-interacting vacuum state on $\Vf$. It is
given by linear extension of the functional 
\be \label{e.3.12} 
\omega_0(\psi(\delta m) W(m)) \coloneqq
\begin{cases}
  e^{\, (1/2) \langle m, D_+ m \rangle} & \text{if} \quad \delta m = 0 \\
    0  & \text{if} \quad \delta m \neq 0 \, , 
\end{cases}
\ee
where the Pauli-Jordan distribution $D$ in equation \eqref{e.3.3}
has been replaced by its positive frequency part $D_+$.
That $\omega_0$ is a positive
functional is a consequence of the fact that
$\omega_0(S(\delta m_j)^*  S(\delta m_k)) = 0$ if
$\delta m_j \neq \delta m_k$. Thus 
\begin{align} \label{e.3.13}
& \omega_0(\big( \sum_j S(\delta m_j) \big)^*
 \big( \sum_k  S(\delta m_k \big) )  \nonumber \\
& = \sum_j \omega_0(S(\delta m_j)^* S(\delta m_j)) 
  = \sum_j \omega_0( \, \big| \,
 \sum_l c_{j,l} \eta_{j,l} V(g_{j,l}) \, \big|^2 \, ) \geq 0  \, .
\end{align}
This lower bound obtains since 
$\omega_0$ is a state on $\Vf$, proving that its extension
to $\Zf$ is a state as well. Moreover, it follows from this
relation that the (non-separable) Hilbert space, which arises
by the GNS-representation
from the functional $\omega_0$ on $\Zf$, decomposes
into sectors labeled by $\delta m$. These sectors
are stable under the action of $\Vf$.

\medskip 
It is also clear from relation
\eqref{e.3.12} that $\omega_0$ is invariant under
Poincar\'e transformations. Hence these transformations
are unitarily implemented in the GNS-representation of $\Zf$. 
But the resulting unitaries
do not depend continuously on these transformations.
This is a consequence of the fact that we have chosen
in definition \eqref{e.3.12} as state on the matter fields 
the singular counting measure. Since we are
primarily interested in the properties of the electromagnetic
field in the presence of external charges,
we do not need to examine here the question of whether
there are more regular extensions of the vacuum state on $\Vf$
to the algebra $\Zf$. 

\medskip
We turn now to the analysis of the states on $\Vf$ which are
obtained by the adjoint action of the operators
$\psi(\delta m) W(m)$ on the vacuum state $\omega_0$.
Since the matter field commutes with all elements of $\Vf$,
the states are given by, cf.\ equations~\eqref{e.2.14} and
\eqref{e.2.10}, 
\be \label{e.3.14}
\omega_m(V(g)) \coloneqq \omega_0(\beta_m(V(g)) = e^{i \varphi_m(g)} 
\omega_0(V(g)) \, , \quad g \in \Cc_1(\RR^4) \, .
\ee
Considering exponentials of the current, \ie $g = \delta d h$
with $h \in \Dc_1(\RR^4)$, definition~\eqref{e.2.9} yields  
\be \label{e.3.15}
\varphi_m(\delta d h) = - \int dx dy \,
(\delta m)(x) D(x-y) (\delta h)(y) \, . 
\ee
Thus, unless $\delta m = 0$, the functional does not vanish for 
suitable test functions~$h$. On the other hand one has
$\omega_0(V(\delta d h)) = 1$ for any choice of $h$. It therefore  
follows from equation \eqref{e.3.14} that the states
$\omega_m$ and $\omega_0$ on $\Vf$ are disjoint 
on regions of Minkowski space where $D \delta m$
is different from $0$. The fact that they are
disjoint can also be seen
by noticing that both states are pure and that 
$\beta_m$ acts non-trivially on the central
elements of $\Vf$.

\medskip
Turning to the energetic properties of the states $\omega_m$,
$m \in \Dc_1(\RR^4)$, 
we make use of the fact that the GNS-representation of
$\Vf$, induced by $\omega_m$, is equivalent to the representation 
on the vacuum Hilbert given by 
\be  \label{e.3.16}
\beta_m(V(g)) \coloneqq
e^{i \varphi_m(g)} V(g) \, , \quad  g \in \Cc_1(\RR^4) \, .
\ee
Assuming for a moment that the time translations on $\Vf$ act 
in these representations by unitary operators $e^{itH_m}$, $t \in \RR$,
one would have that 
\begin{align}  \label{e.3.17}
\ad{e^{itH_m}} & (\beta_m(V(g)) = \beta_m(V(g_t)) \nonumber \\
& = e^{i \varphi_{m_{-t}}(g)} \ad{e^{itH_0}}(V(g))
= \ad{e^{itH_0}}(\beta_{m_{-t}}(V(g)) \, .
\end{align}
Thus the representations given by $\beta_m$ and $\beta_{m_{-t}}$ would
necessarily be equivalent. Since $m$ and $m_{-t}$ have different
supports, this is impossible by the preceding remarks
if $\delta m \neq 0$, \ie in the presence of a non-trivial
charge distribution. That result
is also expected on physical grounds: shifting the external charges
by operations involving only the electromagnetic field requires 
an ill-defined amount of energy. 
In order to cope with this problem, there are two possible strategies.

\medskip \noindent  
(I) \ One may focus on the energy carried by the
transversal part of the electromagnetic field
(the photons) and
ignore the energy needed to shift the  elements
of the center of $\Vf$, such as the current \eqref{e.2.5}. 
This is accomplished by relying on the operator
determining the energy density  
in the vacuum representation of $\Vf$. It is given by the
(normal ordered) operator
\mbox{$(1/2): \! \! \bEi^2 + \bBi^2 \! \! :  \, ,$} where
$\bEi^j \coloneqq F^{0j}$ and
$\bBi^j \coloneqq (1/2) \epsilon_{jkl} F^{kl}$ are the
components of the 
electric, respectively magnetic, field, 
$j,k,l = 1,2,3$. This density is a local 
observable which can be used in any representation of~$\Vf$ in
order to determine the energy content of the transversal
part of the electromagnetic field. The density commutes with
the elements of the center of~$\Vf$ and thus is insensitive to 
their energetic properties. 

\medskip \noindent
(II) \ Alternatively, one can enlarge the universal algebra $\Vf$ to a
gauge invariant algebra with non-trivial center. This is accomplished by
fixing a Lorentz frame and proceeding to the algebra $\Wf_0$ that
is generated by the exponentials of $(\bAi - {\bnabla} \, \xi)$.
These operators consist of 
the spatial components of the vector potential, amended by the
generator~$\xi$ of the matter field, \ie 
\mbox{$\psi(\rho) = e^{i \, \xi(\rho)}$}. 
Putting $\bEi \coloneqq - (\dot{\bAi} -  {\bnabla} \, \dot{\xi})$
and $\bBi \coloneqq {\bnabla} \times (\bAi -  {\bnabla} \, \xi)$,
it is apparent that $\bEi$ and~$\bBi$ satisfy the homogeneous
Maxwell equations, hence they also generate a concrete
version of the universal algebra $\Vf \subset \Wf_0$.
It follows from
relation~\eqref{e.3.7} and the commutativity of the
matter fields that the three components of 
$(\bAi - {\bnabla} \, \xi) $ 
satisfy canonical commutation relations with their time
derivatives. Apart from achieving 
gauge invariance, the matter field~$\xi$ has no
further algebraic impact on the electromagnetic field and
potential. We can therefore 
proceed from $(\bAi - {\bnabla} \, \xi) $ to~$\bAi$ 
again in the following discussion. 

\medskip 
Disregarding the action of Lorentz transformations, the fields 
$\bAi$ and $\dot{\bAi}$ may be regarded as the canonical data of 
three scalar massless fields, satisfying the
wave equation. It is clear then that there exists a
vacuum state on~$\Wf_0$. The resulting time translations
on  $\Wf_0$ preserve the subalgebra $\Vf \subset \Wf_0$,
generated by $\bEi = - \dot{\bAi}$ and $\bBi = \bnabla \times \bAi$. 
It acts covariantly on the current, which    
is no longer an element of the
center of $\Vf$, however. As a matter of fact, a covariant
action on an abelian algebra  would not be compatible
with the existence of a positive generator of the time 
translations according to the Borchers-Arveson theorem
\cite[Thm.\ 3.2.46]{BrRo}.
In the present case, this generator 
is obtained by integration of the energy density  
$(1/2) : \! \! \bEi^2 + \sum_k {\bnabla} A_k \,
{\bnabla} A_k \! \! :$ at fixed time over space.
With this input, one can then study the
dynamics and energy of both, the electromagnetic field 
and the external charges, which are
created by the automorphisms~$\beta_m$.

\medskip
We discuss in the following both strategies. Turning to approach (I),
we make use of the normal ordering procedure in Minkowski
space, relying on the
local fields, \ie we~put
\begin{align}  \label{e.3.18}
& : \!  \bEi^2 + \bBi^2  \! :(x)  
\coloneqq \lim_{\varepsilon}
\big(\bEi(x + \varepsilon) \bEi(x - \varepsilon) 
+ \bBi(x + \varepsilon) \bBi(x - \varepsilon) \nonumber \\
& -  \omega_0(\bEi(x + \varepsilon) \bEi(x - \varepsilon)
+ \bBi(x + \varepsilon) \bBi(x - \varepsilon) \big)
{\mathbf 1} \, .
\end{align}
Here $\varepsilon$ is a
suitable sequence of spacelike translations, tending to $0$.
In any state, where this sequence converges to an operator-valued
distribution, it defines its electromagnetic energy density
relative to the vacuum representation.
Proceeding to the representation $\beta_m$ of $\Vf$ on
the vacuum Hilbert space, we obtain for the  
energy density after a straightforward computation 
\begin{align}  \label{e.3.19}
& \beta_m\big((1/2) : \!  \bEi^2 + \bBi^2  \! :(x) \big) \nonumber \\ 
  & = (1/2) : \!   \bEi^2 + \bBi^2  \! :(x)  +
   \bEi(x) \big( \partial_0
  \underline{\bmi} - {\bnabla} \underline{m}_0 \big)(x) +
   \bBi(x) \big({\bnabla} \times \underline{\bmi})(x) \nonumber \\
  & \, +  (1/2) \big( ( \partial_0 \underline{\bmi}
  - {\bnabla} \underline{m}_0 \big)^2
    +  ({\bnabla} \times \underline{\bmi})^2 \big)(x) \, {\bm 1} \, . 
\end{align}
Here $ \underline{m} \coloneqq D \, m$, thus  $ \underline{m}$
is a solution of the wave equation, and 
$\underline{m}_0$ and $\underline{\bmi}$ are its time and spatial components.
The spatial integral of this density at any
given time~$t$ is well defined as a
quadratic form. Making use of the fact
that on the vacuum Hilbert space one has
${\bdelta} \bEi = 0$ and ${\bnabla} \times \bBi = \partial_0 \bEi$,
one obtains by partial integration 
\begin{align}  \label{e.3.20}
H_m(t) & \coloneqq (1/2) \int \! d\bxi \,
\beta_m\big(: \!  \bEi^2 + \bBi^2  \! :(t, \bxi) \big) \nonumber \\
& = H_0 + \int \! d\bxi \, \big(\bEi \, \partial_0 \underline{\bmi}
- (\partial_0 \bEi) \, \underline{\bmi}) \big)(t, \bxi)  \nonumber \\
& + (1/2) \int \! d\bxi \,
 \big( ( \partial_0 \underline{\bmi}
  - {\bnabla} \underline{m}_0 \big)^2
    +  ({\bnabla} \times \underline{\bmi})^2 \big)(t, \bxi) \, {\bm 1}
\, . 
\end{align}
Since both, $\bEi$ and $\underline{\bmi}$,
are solutions of the wave equation,
the term in the second line does not depend on $t$. Thus the
time dependence of the Hamiltonian $H_m(t)$ is completely contained 
in the c-number contribution. It describes the
mean energy of the perturbed vacuum state. Apart from
this term, $H_m(t)$ is constant in time, so 
it induces a time independent (autonomous)
dynamics of the transversal components of the
electromagnetic field. But it leaves the current invariant 
(being an element of the trivially represented center
of $\beta_m(\Vf))$, hence the obstructions arising
from relation \eqref{e.3.17} do not come into play. 

\medskip
Dynamics of this type have been thoroughly discussed
in the literature by functional analytic methods, cf.\ for
example \cite{Ro} and the discussion in Sect.~4.
In the present algebraic setting their properties follow more easily
from the representation of the algebra $\Zf$, considered
above. Restricting the algebra $\Vf$, acting on the non-separable
representation space, to the sector corresponding to 
$\delta m = 0$, one obtains the standard irreducible vacuum representation
of the electromagnetic field. There the time translations
are implemented by a continuous unitary group
\mbox{$t \mapsto U_0(t)$} with positive
generator. By the adjoint action of the unitary operators
$W(m) \psi(\delta m)$ on this group one obtains continuous unitary 
representations $t \mapsto U_m(t)$ of the time translations
on the algebras $\beta_m(\Vf)$ in the sectors attached to $\delta m$, 
$m \in \Dc_1(\RR^4)$. 
Their generators coincide with the operators~$H_m(t)$
given in equation \eqref{e.3.20}, apart from the c-number term. 

\medskip
Turning to the second approach (II), the corresponding normal-ordered
energy density is defined similarly as in equation \eqref{e.3.18},
where $\omega_0$ is to be replaced by the
extended vacuum functional on $\Wf_0$.
And, similarly as in the preceding discussion, 
we consider the representations $\beta_m$ of $\Wf_0$
on the corresponding vacuum Hilbert space.
In contrast to approach~(I), these representations are now
given by the adjoint action of unitary operators, \ie they are
equivalent to the vacuum representation of~$\Wf_0$. Hence,
the embedding of $\Vf$ into the bigger algebra $\Wf_0$, involving
gauge fields, cures the subtle dependence of its representations
on $\delta m$. 

\medskip
This assertion requires a comment if $m$ has a non-trivial
zero-component $m_0$ since the component 
$A_0$ of the Gupta-Bleuler field is not represented
on the underlying Hilbert space.
The corresponding map $\beta_{m_0}$ does not change $\bAi$, 
but it induces an action on the electric field~$\bEi$
given by
\be \label{e.3.21} 
\beta_{m_0}(e^{i\bEi(\bfi)}) = e^{\, i \int \! dx \, \underline{m}_0(x)
  ({\bm \delta \bm f})(x)} \, e^{i\bEi(\bfi)} \, , \quad
f \in \Dc_1(\RR^4) \, ,
\ee
where $\bfi$ are the spatial components of $f$.
It turns out that in the present representation  
these maps are induced by exponentials
involving only the spatial components of the
vector potential. To see this, 
we make use of the fact that the map $\bgi \mapsto e^{i \bAi(\bgi)}$
is continuous in the strong operator topology 
for~$g \in \Dc_1(\RR^4)$
varying continuously with respect to the single particle seminorm
\be \label{e.3.22} 
\| \bgi \|_0^2
\coloneqq \int \! \frac{d\bpi}{2|\bpi|} \, | \widetilde{\bgi}(|\bpi|,\bpi)|^2
\, , \quad g \in \Dc_1(\RR^4) \,  .
\ee
Note that the kernel of this seminorm consists of functions which
vanish on the zero-mass shell. So the fields, being solutions
of the wave equation, also vanish on these functions. Whence the
unitary operators $e^{i \bAi(\bgi)}$
can continuously be extended to all real functions in
the single particle space. After a moment of reflection, it follows that the
(real, vector-valued) function
\be  \label{e.3.23} 
x \mapsto \bni(x) \coloneqq {\bnabla} \, \!
\partial_0 {\bm \Delta}^{-1} m_0(x)
\ee
is contained in this space and the adjoint action of
the unitaries $e^{i \bAi(\bni)}$ on $\Wf_0$ coincides with the action of
$\beta_{m_0}$. Thus all automorphisms
$\beta_m$, $m \in \Dc_1(\RR^4)$, are unitarily
implemented in the vacuum representation of $\Wf_0$, as claimed. 

\medskip
In view of this result it is clear from the outset that the time translations
are unitarily implemented in all representations $\beta_m$ of $\Wf_0$
and their generators are bounded from below. 
Nevertheless, it is of interest to have a look at the corresponding
Hamiltonians and to compare them with the results in (I). This is 
accomplished by noting that the unperturbed energy density,
given above, can
be recasted, disregarding partial derivatives of local operators which 
vanish upon integration over all space. One has, up to
such negligible terms,  
\be \label{e.3.24}
(1/2) : \! \! \bEi^2 + \sum_k {\bnabla} A_k \,
  {\bnabla} A_k \! \! :(x)  \ = 
(1/2) :  \! \! \bEi^2 + \bBi^2 + ({\bm \delta \bAi})^2 \! \! :(x) \, .  
\ee
Comparing this with (I), it is apparent that 
the energy density related to proper time translations of
the current is encoded in the last term. The spatial
integral of this density yields the Hamiltonian in
the vacuum representation of~$\Wf_0$. The Hamiltonians
in the representations $\beta_m$ are obtained from it by
the adjoint action of the unitaries $e^{i \bAi(\bni - \bmi)}$,
where $\bmi$ are the spatial components of $m$ and
$\bni$ was defined in equation~\eqref{e.3.23}. In
particular, the mean energy of the perturbed vacuum
states is in general bigger than the proportion of the
transversal electromagnetic field, determined in (I).
We refrain from presenting here these straightforward computations. 

\medskip
Let us point out in conclusion that the case of external charges,
considered here in the Gupta-Bleuler formalism, can also be studied in other
approaches to the treatment of gauge fields, which inevitably
accompany the charges. The Gupta-Bleuler formulation
has the advantage that the localization properties of
charge-containing operators can be
described in a simple, transparent manner. It leads to a completely local
formulation in case of states with vanishing global charge. In the
subsequent section we will show how states carrying a non-vanishing
global charge can be approximated by these neutral states,  
thereby developing the long-range localization,
known from physical gauges. 

\section{Passage to charged states}
\label{sec4}
\setcounter{equation}{0}

Using the framework of the preceding Sect.\ 3, we consider now  
sequences of neutral states which describe bi-localized external
charge distributions. 
Thereby, one of these 
distributions is kept fixed and
the other (compensating) charge distribution
is moved to spacelike infinity. In this way, 
the influence of the compensating charges on local observables
is suppressed and the resulting limit states are charged.
We adopt in this analysis arguments given in \cite{BuCiRuVa2021},
where we need to appropriately
increase the localization regions occupied by
the compensating charges in order to maintain control on the
energetic properties of the states. As a result, the charged
limit states typically differ from the vacuum state in spacelike cones,
where fields with an asymptotic behavior of Coulomb-type appear.  

\medskip
Proceeding to the construction, the essential step consists of the proof of 
existence of suitable functions $m \in \Dc_1(\RR^4)$ that
enter in the automorphisms~$\beta_m$. We begin with the
distributions, where $a,b \in \RR^4$, 
\be \label{e.4.1}
x \mapsto m^\mu(a,b)(x) \coloneqq (b - a)^\mu \int_0^1 \! du \,  
\delta(u b + (1-u) a - x) \, .
\ee
They satisfy the equation 
$\partial_\mu m^\mu(a,b)(x) = \delta(x-a) - \delta(x-b)$,
where $\delta$ is the four-dimensional Dirac measure.
Integrating $m^\mu(a,b)$ with test functions depending on $a, b$ 
and having compact support, 
one obtains a test function with regard to $x$ whose support is
contained in the convex hull of all pairs of points $a,b$ in
the support of the 
chosen functions. We pick now a smooth charge distribution
$a \mapsto \vartheta(a)$
with compact support, which is kept fixed,  and real test functions 
$b \mapsto \sigma(b)$ whose supports will
be moved to spacelike infinity, while their integrals
are kept equal to $1$. Putting
\be \label{e.4.2} 
x \mapsto m(\sigma)^\mu(x) \coloneqq
\int \! da db \, \vartheta(a) \sigma(b-a) \, m^\mu(a,b)(x) \, ,
\ee
equation \eqref{e.4.1} implies
\be \label{e.4.3} 
\partial_\mu  m(\sigma)^\mu(x) = \vartheta(x) -
 \vartheta * \sigma(x) \, . 
\ee
In the situation of interest here, these  distributions occupy  
two spacelike separated regions, where both components
have fixed spacetime integrals 
with opposite signs. So they compensate each other. 

\medskip
The charge content in the regions can be determined by the
current~\eqref{e.2.5} with test functions of the form
$x \mapsto h(x) \coloneqq \tau(x_0) \chi(\bxi)$. Here one chooses 
smooth characteristic functions  $\chi$ of the regions in $\RR^3$ 
in which the charge is to be determined and test functions $\tau$ with
support in a small time interval about the time where this is to happen;
the integrals of the latter functions are to be equal to $1$.

\medskip 
Let $\Oc_h$ be the  causal completion of the convex hull of 
$\text{supp} \, h$. It is the spacetime region where the operation of charge
measurement takes place. Due to the unavoidable choice of
smoothed characteristic
functions $\chi$ and Dirac measures~$\tau$, the charge measurements 
become effective only in certain specific subregions of $\Oc_h$.
These subregions come close to the surrounding region~$\Oc_h$ by a suitable
choice of $\chi$ and $\tau$, 
cf.~\cite{BuCiRuVa2021}. Whenever a region
$\Oc_0 \subset \RR^4$ is con\-tained in such a subregion, 
we write \mbox{$\Oc_0 \Subset \Oc_h$}. By arguments given in the
proof of \cite[Lem.~2.2]{BuCiRuVa2021}, one obtains 
for the functionals \eqref{e.2.9}, depending on
the given charge distribution \eqref{e.4.3}, 
\be \label{e.4.4}
\varphi_{m(\sigma)}(h) =
\begin{cases}
  \int \! dx \, \vartheta(x) & \text{if} \ \ 
  \text{supp} \, \vartheta \Subset \Oc_h \, ,
  \ \text{supp} \, \vartheta * \sigma \perp \Oc_h  \\
  0 & \text{if} \ \ \text{supp} \, \vartheta \cup
  \text{supp} \, \vartheta * \sigma \Subset   \Oc_h \, ,   \\
  -  \int \! dx \, \vartheta(x) & \text{if} \ \
  \text{supp} \, \vartheta * \sigma \Subset \Oc_h \, ,
   \ \text{supp} \, \vartheta  \perp \Oc_h
\end{cases}     
\ee
These relations are a distinctive consequence of Gauss's law. 
Since the regions $\Oc_h$ are bounded, it is apparent that by moving  
the support of $\sigma$ to spacelike infinity, only
the charge in $\text{supp} \, \vartheta$ remains visible in the limit. 
Thus, composing the vacuum functional with the automorphisms
$\beta_{m(\sigma)}$, the resulting limits describe charged
states. That these limit states exist for suitable
sequences of $\sigma$ can be seen as follows. 

\medskip
Picking any test function $\sigma$ with compact support in the spacelike
complement of the origin of $\RR^4$ and integral $1$, one proceeds to the
scaled functions $x \mapsto \sigma_r(x) \coloneqq (1/r^4) \, \sigma(x/r)$
for $r > 0$. The support of these functions approaches spacelike
infinity in the limit of large $r$ and the resulting 
functionals $\varphi_{m(\sigma_r)}$ converge on $\Dc_1(\RR^4)$
in this limit. For the proof one 
needs to determine the Fourier transforms of $m(\sigma_r)$.
Putting $n \coloneqq (1/|p|) \, p$, where $|p|$ is the Euclidean length
of $p$, and $b \mapsto \sigma^\mu(b) \coloneqq \sigma(b)
\hspace{0.3pt}  b^\mu$,
they are given by
\begin{align} \label{e.4.5}
p \mapsto \widetilde{m}(\sigma_r)^\mu(p) & =
\widetilde{\vartheta}(p) \int \! db \, \sigma^\mu(b) \, 
          {\frac{e^{i rbp} -1}{ibp}} \nonumber \\
 & = (2 \pi)^2 \, \widetilde{\vartheta}(p) \, |p|^{-1} \! \int_0^{r|p|} \! du \,
          \widetilde{\sigma}^\mu(u \hspace{0.3pt} n) \, .
\end{align}
Thus one obtains for $p \neq 0$ in the limit of large $r$
the singular function
\be \label{e.4.6} 
p \mapsto  \widetilde{m}(\sigma_\infty)^\mu(p) =
(2\pi)^2 \, \widetilde{\vartheta}(p) \, |p|^{-1} \varrho^\mu(n) \, ,
\ee
where $n \mapsto \varrho^\mu(n) \coloneqq \int_0^\infty \! du \,
\widetilde{\sigma}^\mu(u \hspace{0.3pt} n)$ is continuous and bounded. 
By a routine computation this yields for $f \in \Dc_1(\RR^4)$ 
\begin{align} \label{e.4.7} 
& \varphi_{m(\sigma_\infty)}(f) 
  \coloneqq - \lim_{r \rightarrow \infty} \int \! dx dy \,
  m(\sigma_r)^\mu(x) \, D(x-y) \, f_\mu(y) 
\nonumber \\  
& = - i (2 \pi)^3 \!  \int \! dp \, \epsilon(p_0)  \delta(p^2) \, 
\tilde{\vartheta}(p) |p|^{-1} \! \varrho^\mu(n) \, 
\widetilde{f}_\mu(-p) \nonumber \\
& = (2 \pi)^3 \, \text{Im} 
\Big(\int \! \frac{d\bpi}{ \sqrt{2} \, |\bpi|^2} \,   
{\widetilde{\vartheta}(\underline{p})
\varrho^\mu(\underline{n})} \widetilde{f}_\mu(-\underline{p}) \Big)
\, .
\end{align}
Here we have put $\underline{p} \coloneqq (| \bpi |, \bpi)$ and
$\underline{n} \coloneqq (1/\sqrt{2}) (1, \bpi / | \bpi |)$. The
singularity in the integral on the last line
is absolutely integrable, so the limit functionals are
well-defined on $\Dc_1(\RR^4)$. 

\medskip
It follows that the automorphisms $\beta_{m(\sigma_r)}$
converge pointwise on $\Vf$ in the strong operator topology
for asymptotic $r$ on the vacuum Hilbert space.
The states appearing in the resulting limit
representation carry the global charge $\int \! dx \, \vartheta(x)$. 
Yet the underlying unitary operators
$W(m(\sigma_r)) \psi(\delta m(\sigma_r))$, which implement
the automorphisms, 
do not have a meaningful limit. Thus, in order to
determine the energetic properties
of the limit states, one cannot proceed as in
Sect.\ 3. Instead, one has to rely on functional analytic methods. 

\medskip 
We consider first the electromagnetic energy content of the limit
states, as described in approach (I) in Sect.\ 3. 
In doing so, we make use of the fact that the middle term in the Hamiltonian
\eqref{e.3.20} remains unchanged if one replaces the spatial components
$\bmi$ of $m$ by 
\be \label{e.4.8}
x \mapsto \bgi_\bmi(x) \coloneqq
(\bmi - \bnabla \bDelta^{-1} \bdelta \bmi)(x) \, .
\ee
The gradient term does not contribute to the integral as one sees
by partial integration, making use of the fact that $\bdelta \bEi = 0$
in the vacuum representation of $\Vf$.
The functions $\bgi_\bmi$ are smooth and satisfy 
$\bdelta \bgi_\bmi = 0$. 
Moreover, their seminorm~\eqref{e.3.22} is finite. 
So the unitary operators $V(g)$,  $g \in \Cc_1(\RR^4)$,
can continuously be extended to unitary operators~$V(\bgi_\bmi)$
in the strong operator topology on the vacuum Hilbert space.
By a routine computation, one finds that the adjoint action
of~$V(\bgi_\bmi)$ on the density \eqref{e.3.8} yields  
the Hamiltonian~\eqref{e.3.18} after integration,
disregarding again the c-number term.

\medskip
Given any sequence $m(\sigma_r)$, $r > 0$, 
these observations put us into the position to use the results
in \cite{Ro} for the determination of the electromagnetic
energy content of the limit states. Whereas the functions $\bgi_{\bmi(\sigma_r)}$ 
do not converge for large $r$ with regard to
the seminorm \eqref{e.3.22}, they have limits with regard to
its mollified version,  
\be \label{e.4.9} 
\| \bgi_{\bmi(\sigma_r)} \|_1^2
\coloneqq \int \!
\frac{d\bpi}{2|\bpi|} \, \Big( \frac{\bpi^2}{1 + \bpi^2} \Big)^{1/2} \, 
| \widetilde{\bgi}_{\bmi(\sigma_r)}(|\bpi|,\bpi)|^2 \, .
\ee
(The kernel of this seminorm likewise consists of functions
that do not contribute in the fields.)  
Due to the mollifying factor in the integral \eqref{e.4.9}, 
the singularity of the limit
function $\widetilde{\bgi}_{\bmi(\sigma_\infty)}$ at zero
momentum is tamed, cf.\ equation~\eqref{e.4.6}. In fact, the sequence
$\widetilde{\bgi}_{\bmi(\sigma_r)}$, $r > 0$, converges with respect to
the mollified seminorm to
$\widetilde{\bgi}_{\bmi(\sigma_\infty)}$.

\medskip 
Making use of 
arguments in \mbox{\cite[Prop.\ 3]{Ro}}, it follows that the
energy densities, obtained by applying the automorphism $\beta_{m(\sigma_r)}$
to the vacuum density~\eqref{e.3.18}, determine 
by integration over all space a meaningful limit dynamics. 
In more detail: let $H_0$ be the Hamiltonian in the vacuum representation
of~$\Vf$. Because of the convergence properties of
the functions~$\bgi_{\bmi(\sigma_r)}$ for asymptotic $r$, the 
sequence of cocycles
\begin{align} \label{e.4.10}
  &  t \mapsto V(\bgi_{\bmi(\sigma_r)}) \, e^{itH_0}  V(\bgi_{\bmi(\sigma_r)})^*
     e^{-itH_0} = V(\bgi_{\bmi(\sigma_r)}) V(\bgi_{\bmi(\sigma_r), t})^*
  \nonumber \\ 
&  = e^{-(i/2) \langle \bgi_{\bmi(\sigma_r)} , \, D \, (\bgi_{\bmi(\sigma_r)}
  - \bgi_{\bmi(\sigma_r), t}) \rangle}
  \, V(\bgi_{\bmi(\sigma_r)}  - \bgi_{\bmi(\sigma_r), t})
\end{align}  
converges in this limit in the strong operator topology.
Note that the Fourier transforms of the
differences $(\bgi_{\bmi(\sigma_r)}  - \bgi_{\bmi(\sigma_r), t})$
vanish at zero momentum  and 
converge for large $r$ with regard to the single particle seminorm
\eqref{e.3.22}, uniformly on compact subsets of $t$.
These facts imply that the limit dynamics exists in
approach~(I). It is given as a limit in the strong operator topology 
\begin{align} \label{e.4.11}
t \mapsto e^{i t H_{ (I) }(m(\sigma_\infty))} &  \coloneqq
\lim_{r \rightarrow \infty} V(\bgi_{\bmi(\sigma_r)}) \, e^{it H_0} \,
V(\bgi_{\bmi(\sigma_r)})^* \nonumber \\ 
& \hspace*{2.5pt} =
\lim_{r \rightarrow \infty} V(\bgi_{\bmi(\sigma_r)}) V(\bgi_{\bmi(\sigma_r), t})^*
\, e^{it H_0} \, .
\end{align}
The limit dynamics is 
strongly continuous, has a positive generator, but   
leaves the center of $\beta_{\bmi(\sigma_\infty)}(\Vf)$
pointwise fixed. So it does not act
properly on the charge operators. But it fully describes 
the dynamics and energy content of the
transversal components of the electromagnetic field 
in the charged sectors. 

\medskip
In a similar manner one proceeds in approach (II). There
the dynamics in the vacuum sector of $\Wf_0$ 
leaves $\Vf \subset \Wf_0$ invariant and
acts covariantly on the current. The representations in
the locally charged sectors are obtained by acting on $\Wf_0$ in the
vacuum sector with the automorphisms $\beta_m$.
One then chooses sequences of functions $m(\sigma_r)$,
describing as above bi-localized charge distributions, 
and considers the resulting automorphisms  $\beta_{m(\sigma_r)}$,  \mbox{$r > 0$}.
As was explained in Sect.\ 3, these automorphisms are 
implemented on the vacuum Hilbert space by unitary operators
$e^{i \bAi(\bni(\sigma_r) - \bmi(\sigma_r))}$, $r > 0$.  Here
$\bmi(\sigma_r)$ denotes again the spatial components of $m(\sigma_r)$, and
$\bni(\sigma_r)$ is determined by the time component $m_0(\sigma_r)$
according to equation \eqref{e.3.23}.

\medskip 
One then determines the Fourier transforms of 
$(\bni(\sigma_r) - \bmi(\sigma_r))$ and finds that these sequences  
have similar convergence properties
as those of the sequences $\bgi_{\bmi(\sigma_r)}$ 
in approach~(I), $r > 0$. They also develop a singularity
at zero momentum in the limit whose restriction to the zero-mass shell 
has the same form as in equation \eqref{e.4.6}. 
Thus, by applying the preceding 
arguments, one arrives at the conclusion that the limit dynamics exists
in approach~(II) in all charged sectors. It is given~by 
\be \label{e.4.12}
t \mapsto e^{i t H_{ (I \! I) }(m(\sigma_\infty))}   \coloneqq
\lim_{r \rightarrow \infty} e^{i \bAi(\bni(\sigma_r) - \bmi(\sigma_r))} \,
e^{itH_0} \, e^{-i \bAi(\bni(\sigma_r) - \bmi(\sigma_r))} \, ,
\ee
where $H_0$ is the Hamiltonian in the vacuum
representation of $\Wf_0$. The resulting unitary group
is continuous and has a positive generator. 
We omit the proof since it completely
parallels the preceding discussion. 

\medskip
Let us finally mention that the globally charged representations, obtained
in this manner, are locally normal relative to the vacuum representation.
Yet, depending on the choice of the approximating functions
$m(\sigma_r)$, $r>0$, there exists
an abundance of limit states that carry the same global charge, 
but are mutually disjoint, nevertheless.
These states differ by the asymptotic configurations of the
electromagnetic field or, alluding to the particle picture, by infinite
clouds of low energy photons. Since these ``infrared problems'' are
widely known, we do not dwell here on this issue any further.

\section{Conclusions}
\label{sec5}
\setcounter{equation}{0}

Proceeding from the universal algebra of the
electromagnetic field, providing a general framework for the
discussion of electromagnetism, we have studied the
question of whether the presence of gauge fields can be
uncovered from the gauge invariant observables.
All basic features of the electromagnetic field are
encoded in this algebra, in particular the
homogeneous Maxwell equations, the existence of a
current operator, and the
condition of locality. These features are conveniently
expressed in a C*-algebraic framework, from which the
underlying unbounded field operators can be recovered
in regular representations. In order to uncover in this 
setting possible traces of gauge fields, we have studied
the impact of compactly localized external charge distributions
on the algebra; they give rise to a vanishing global charge.
These charge distributions can be traced and analyzed by means of 
the current operator in the algebra,
making use of Gauss's law. 

\medskip 
It turned out that the external charge distributions   
induce outer automorphisms of the universal algebra. 
This fact reveals the presence of local degrees of
freedom, which are not incorporated in the algebra.
In order to represent the automorphisms
by the adjoint action of unitary operators,
describing these additional degrees of freedom, the universal
algebra had to be extended. Yet 
this first extension did not unambiguously describe the
unitaries: they can  
be modified by an abundance of local transformations
without changing their action on the observables.
These transformations can be interpreted as local gauge
transformations, whereby the generating elements
of the extended algebra correspond to the exponentials
of gauge fields.

\medskip
In order to completely fix the unitary implementers (up to phase
factors), one has to introduce further operators, which can 
be interpreted as charged matter fields. They may be of  a 
bosonic, fermionic, or classical nature. In the present case of
external charges, we considered the simple case of classical
charged matter. It affects the
quantized electromagnetic field only
through the accompanying tail of gauge fields. 
The gauge invariant combinations of gauge and matter
fields define a physically meaningful extension of the universal algebra. 
It is generated by local, Poincar\'e covariant implementers
of the charge-containing automorphisms. The appearance of
gauge and matter fields is thus encoded in the structure of the
universal algebra and manifests itself already
in the simple case of external charges.

\medskip
We have illustrated this abstract framework by representing the
gauge and matter fields concretely in a setting of
Gupta-Bleuler type. There the universal
algebra is canonically embedded in the Weyl algebra generated
by the Gupta-Bleuler fields. Moreover,  
the framework is enriched by some dynamical input,
deriving from the underlying wave equation.
In this formalism the current does not vanish,
but it is affiliated with the center of the universal algebra 
as a consequence of the dynamical input. 
Since the Gupta-Bleuler fields incorporate operators which
induce gauge transformations, we had also 
to impose for the sake of consistency commutation relations
between the Gupta-Bleuler and matter fields. To the best of our
knowledge, these commutation relations have not been considered
before in the literature. 

\medskip
Applying the implementers of the
charge-containing automorphisms to the non-interacting vacuum
state of the universal algebra, one obtains further
irreducible representations of this algebra.
They are mutually disjoint, even locally,
for different choices of charge distributions. Nevertheless,
there exist energy operators in these representations that are
bounded from below and induce a dynamics on the transversal degrees of
freedom of the electromagnetic field (the photons). 
Let us mention as an aside that these dynamics, together with their 
spectral properties, exist in all Lorentz frames. In other words,
the spacetime translations are unitarily implemented in the
representations and satisfy the relativistic spectrum condition.
However, they act trivially on the current operator which,
being an element of the center, is represented by multiples
of the identity. 

\medskip
It is clear from the outset that one cannot have a dynamics
with positive generator that acts non-trivially on an abelian algebra.
We have therefore enlarged the concretely given 
universal algebra to ensure a covariant action of the dynamics 
on the current. This enlargement was accomplished by 
adding to the algebra the spatial components of the Gupta-Bleuler
field. The resulting algebra complies again with the defining
properties (axioms) of the universal algebra. As a result of this step, 
all representations, obtained by applying the
charge-generating unitary operators to the vacuum state of the
extended algebra, become unitarily equivalent. Moreover, the
resulting dynamics has a positive generator,
it leaves the universal algebra invariant, and it acts covariantly
on the current. Again, this property obtains in all Lorentz
frames. 

\medskip
For completeness, we have also discussed the passage from 
neutral states to states carrying a global charge. This
was accomplished
by the well-known method to create pairs of compensating
charges and to move one of them to spacelike infinity. Just as 
for localized charge distributions, the resulting
representations of the universal algebra
have a dynamics with positive generator. In case of the
extended algebra, the charged representations are also 
locally normal with respect to each other. But, for given
value of the global charge, they are in general disjoint, remembering  
the localization properties of the approximating states.
This feature is in accordance with the poor localization properties 
of charged states, known from physical gauges. It is the
origin of the notorious infrared problems in quantum electrodynamics. 

\medskip 
In summary, starting from the local gauge invariant
observables, we have established the occurrence of gauge fields
as an inherent feature of the quantized electromagnetic field.
First of all, gauge fields inevitably accompany 
operations that generate local and global charge distributions.  
They must be amended by local charged matter fields
which compensate the gauge charges carried by the
gauge fields. Secondly, the gauge fields
also enter in a fundamental way into the description of the 
dynamics. They are needed to describe the temporal evolution of the
current operator, which is essential for the
verification of Gauss's law.
Thus, whereas the gauge fields cannot be observed directly,
their impact on observables can be determined by observations. 
In this way, they become an integral part of physics.

\section*{Acknowledgment}

\vspace*{-2mm}
DB  gratefully acknowledges the 
support and hospitality of Roberto Longo and the 
University of Rome Tor Vergata, which made
this collaboration possible. He is also grateful
to Dorothea Bahns and the Mathematics Institute
of the University of G\"ottingen for their continuing hospitality. 
FC and GR acknowledge the MIUR Excellence Department Project awarded to the
Department of Mathematics, University of Rome Tor Vergata, CUP
E83C18000100006, the ERC Advanced Grant 669240 QUEST ``Quantum
Algebraic Structures and Models'' and GNAMPA-INdAM. 
FC was supported in part by MIUR-FARE R16X5RB55W $\! \!$
QUEST-NET ``Operator Algebras
and (non)-equilibrium Thermodynamics in Quantum Field Theory''.

\end{document}